\documentclass[journal]{IEEEtran}

\usepackage{amsthm}
\usepackage{xcolor}
\usepackage{amsmath,amsfonts}
\usepackage{array}
\usepackage[caption=false,font=scriptsize,labelfont=sf,textfont=sf]{subfig}
\usepackage{textcomp}
\usepackage{stfloats}
\usepackage{url}
\usepackage{verbatim}
\usepackage{graphicx}
\usepackage{cite}

\usepackage[ruled, vlined, linesnumbered]{algorithm2e}  

\SetCommentSty{smallcommentsty} 
\usepackage{xifthen}
\usepackage{booktabs}

\usepackage{multirow}

\input{XX-model-defs.sty}

\begin{document}

\title{A Decoupled Approach for Modeling N-k Interdiction with Induced Failures in Electric Power Grids}

\author{Andrew Chio, Kaarthik Sundar, Russell Bent, Nalini Venkatasubramanian%
\thanks{Andrew Chio and Russell Bent are with the Applied Mathematics and Plasma Physics Division, Los Alamos National Laboratory, Los Alamos, NM 87544 USA (emails: achio@lanl.gov, rbent@lanl.gov).}%
\thanks{Kaarthik Sundar is with the Information Systems and Modeling Group, Los Alamos National Laboratory, Los Alamos, NM 87544 USA (e-mail: kaarthik@lanl.gov).}%
\thanks{Nalini Venkatasubramanian is with the Department of Computer Science, University of California, Irvine, Irvine, CA 92697 (email: nalini@uci.edu).}%
}




\IEEEpubid{}

\maketitle

\begin{abstract}
The electric power grid is a critical infrastructure system subject to different types of failures which can have devastating impacts on society. 
One framework to study the resilience of the grid is the $\Nk$ interdiction model, which introduces an attacker and defender model to identify grid vulnerabilities. 
However, traditional $\Nk$ interdiction models characterize the power grid with steady-state assumptions. 
One key aspect missing from this analysis is the behavior of the grid as it evolves from the pre-failure grid state and post-failure grid state. 
During such a transition period, automated protection systems, transient instabilities, and other phenomena can trigger cascading failures that can occur before grid operators can respond.
To bridge this gap, this paper first introduces the generalized induced failure $\Nk$ ($\Nkkp$) problem, where an initial set of $\budget$ failures induces an additional set of $\budget[']$ failures.
In this context, we propose a disjoint optimization approach that decouples the computation of cascading failures from the interdiction model. 
This system response model, furthermore, is treated as a black box module without compromising the underlying performance guarantees of the interdiction model. 
Extensive experiments on several benchmark power grid case studies showcase the efficacy and scalability of the approach.

\end{abstract}


\section{Introduction}
\label{sec:introduction}

The electric grid is a critical infrastructure system that consists of an interconnected network of generators, substations, buses, and transmission/distribution lines, which work to supply power to consumers.
As power moves through the grid, systems for real-time monitoring and control, including Supervisory Control and Data Acquisition (SCADA) and automated protection systems, are used to help maintain safe operation.
However, recent events, e.g., natural disasters and man-made attacks~\cite{watson2019modeling}, have revealed vulnerabilities in the grid, and their severe consequences on cities and communities.

To ruggedize the grid against such failures, research efforts often focus on identifying the core components (generators and lines) that the grid depends on most. 
In this context, one framework used is the $\Nk$ interdiction model, which is based on the zero-sum Stackelberg game~\cite{salmeron2004analysis, donde2008severe, cicilio2020electrical}.
The standard setting considers two opposing players: 
the \emph{attacker} who selects $\budget$ components in a network of size $N$ to simultaneously (or near-simultaneously) fail; 
and the \emph{defender} who controls the grid response to minimize the disruption to service caused by the failure. 
Traditionally, worst-case $\Nk$ models assume that the electric grid before and after failure is fully characterized by a \emph{steady-state model}. 
In practice, these models capture the total amount of load that can be carried through the network while allowing for operator responses, such as generator re-dispatch and transmission switching. 
However, one key aspect missing from $\Nk$ interdiction analysis is the trajectory that captures the evolution of the grid from its pre-failure state to its post-failure state. 
During this transition period, the automated activation of protection systems and transient instabilities may potentially cause additional failures in the network. As a result, there may be no feasible trajectory that allows the grid to reach the desired post-failure state. This can yield steady state analyses that underestimate the disruption of a $\Nk$ event.

One major challenge lies in constructing appropriate models to bridge the disconnect between steady-state assumptions and the transient behavior of the grid.
This is due to additional non-convexities introduced when directly incorporating these features into traditional optimization-based $\Nk$ interdiction models. This can cause the interdiction models to become computationally intractable or produce suboptimal solutions.
To address this, many emerging methods have looked to capture some aspects of the dynamics. 
Initial efforts include the $N \shortminus 1 \shortminus 1$ interdiction model, which studies the evolution of the network when two failures occur one after another, with a short, interleaved recovery period~\cite{fan2012n, vazquez2020reliable}.
Other work has explored limitations on the degree to which generators can re-dispatch~\cite{chio2025sequin}, examining failures under shorter timescales~\cite{rocco2021multi, ren2008long}, and running simulation studies on full grid dynamics~\cite{powerworld,lara2023powersimulationsdynamicsjl,app9050923, PyPSA} for $\Nk$ events.
However, these techniques only begin to tackle the core problem of how to identify the worst-case $\budget$ failures for a network, while accounting for both transient dynamics and steady-state constraints.

In this paper, we develops a model that integrates the standard steady-state assumptions on interdiction, and the immediate system behavior following grid failures. 
We begin by considering an extended $\Nk$ analysis, which we term the $\Nkkp$ problem, where the defender model is augmented with a model for the immediate response of the grid.
In doing so, the defender becomes responsible for computing how the initial $\budget$ failures lead to an additional set of $\budget[']$ failures.
We then leverage the key insight that this computation for the system response can be decoupled from the optimization problem. This allows us to reduce the complexity of the defender while preserving solution optimality. 
Moreover, such a response model can be treated as an abstract black box, which enables 
\clearpage 
\noindent interoperability with a variety of complex computational models for system response~\cite{powerworld, lara2023powersimulationsdynamicsjl}. 
Our key contributions are:

\begin{itemize}
\item The formulation of the $\Nkkp$ problem as a Stackelberg game, where the initial failure of $\budget$ components in the grid induces $\budget[']$ subsequent failures of other components as part of the immediate automated system response. 
\item The development of an approach incorporating arbitrarily 
complex grid response models for $\Nkkp$ analysis \emph{without impacting the underlying performance guarantees}.
\item The design of a disjoint optimization framework that decouples the immediate system response from the attacker and the defender in the interdiction, enabling an efficient solution for the $\Nkkp$ problem. 
A key novelty of this approach is its interoperability with different response models, which allows for flexibility in capturing dynamics at a desired scale and fidelity.
\item An extensive evaluation of the proposed approach on several benchmark grid case studies, which highlights its utility and scalability in studying grid vulnerabilities.
\end{itemize}

We organize the paper as follows:
Section~\ref{sec:related-works} reviews related literature on $\Nk$ interdiction models for contingency analysis, and discuss the state of the art in modeling grid dynamics. 
Section~\ref{sec:preliminaries} then presents the preliminary formulation for the electric grid, and describes the standard $\Nk$ interdiction model. 
We define the new $\Nkkp$ problem in Section~\ref{sec:nkkp-interdiction-problem}.
Then, in Section~\ref{sec:approach}, we present our proposed disjoint approach by constructing it through a progression of steps that build on one another.
Our experiments in Section~\ref{sec:experiments} validate our model, and Section~\ref{sec:conclusion} presents conclusions and future directions. 

\section{Related Literature}
\label{sec:related-works}

We next review related literature for contingency analysis with $\Nk$ interdiction models in power grid systems. 
We then discuss the realism and scalability provided by different types of models, examine techniques used to capture grid dynamics, and place our paper's contributions in the broader literature. 

\textbf{On $\Nk$ contingency analysis}.
Studying grid resilience often involves using contingency analysis to identify and evaluate the reliability risks of potential faults~\cite{willis1997power,chen2005identifying}.
$\Nk$ interdiction models provide a framework to study contingencies through a Stackelberg game.  
Early work by~\cite{salmeron2004analysis} introduced bi-level mathematical models to capture the network interdiction setting for the electric grid. This is used as a basis from which many variants of attacker and defender models have been defined. 
The realism and the scalability of computation depend on several aspects modeling choices in the interdiction.
The standard attacker has full knowledge of the network state and its operation~\cite{fan2011economic, lopez2017assessment, aquino2023comparing}. This conservative assumption allows the attacker to explore worst-case scenarios of grid components to fail in the $\Nk$ contingency. 
However, the power of the attacker to fail certain grid components, such as generators or lines, can vary. While many models allow attackers to disable any component in the grid~\cite{salmeron2009worst, aquino2023comparing}, other efforts have examined different types of constraints to better reflect real-world conditions. Such work has incorporated the probabilistic failure of lines~\cite{sundar2017probabilistic}, scenarios with ruggedized grid components~\cite{wu2016efficient, lai2019tri}, and spatial limitations for enhanced credibility of attacks~\cite{sundar2021credible}.
For the defender, models differ in how they represent the grid operator's capacity to prepare for and respond to failure. This ranges from full observability and control which occurs under normal operating conditions~\cite{salmeron2009worst, sundar2021credible}, to models that look to include uncertainties stemming from factors such as false data injections~\cite{wang2019dynamic, lu2022differential} and unpredictable demand patterns~\cite{rocchetta2018assessment}. 
The defender model can also consider hardening budgets that proactively protect some grid components from failure, as in~\cite{wu2016efficient, mastin2023best}.

The scalability of these methods provides another central challenge. 
The $\Nk$ interdiction problem is combinatorial and NP-hard~\cite{wood1993deterministic, smith2013modern}, so full enumeration is often infeasible beyond small cases. 
Thus, different modeling assumptions consider how to balance the tradeoff between computational efficiency and accuracy of capturing the underlying complexities of power flow. 
This includes efforts such as~\cite{johnson2022scalable,sundar2024exact}, which examine how heuristic bounds on the dual variables of the defender model can be made tighter for the optimization. 
The role of neural networks in capturing network state~\cite{aquino2023comparing}, and network science-based approaches that exploit spatial and empirical grid behavior patterns during failure~\cite{zhou2023most, chio2025sequin} have also been explored.
A key limitation is the lack of modeling for grid dynamics: many works rely on the steady state network representations. As a result, these models can potentially disregard grid behaviors during or immediately after failures. 

\textbf{Towards capturing grid dynamics}.
Integrating grid dynamics for this interdiction problem requires models that include a temporal dimension.
The $N \shortminus 1 \shortminus 1$ model studies the state of the network when a single fault is followed by a short recovery period and then a second fault~\cite{fan2012n, vazquez2020reliable}. 
Extensions have also examined sequential attacks on the grid, where grid components are failed iteratively. This can be due to a sequential attacker~\cite{yan2016q, chio2025sequin}, or through the modeling of shorter timescales~\cite{rocco2021multi}, or cascading failures~\cite{ren2008long}.
Finally, a number of power system simulation libraries have also been developed that focus on simulating the full dynamics of the grid, e.g., simulating waveforms and generator operation. Examples of simulations include PowerWorld~\cite{powerworld}, PowerSimulationsDynamics~\cite{lara2023powersimulationsdynamicsjl}, Mosaik~\cite{app9050923}, and PyPSA~\cite{PyPSA}, among many others.
However, a major limitation of these works is their scalability, they can be used to enumerate all $\Nk$ interdictions for a given $\budget$. Thus, many case studies are small in scale, and depending on the level of detail used to model the operation of the grid, computationally expensive to run. 

\textbf{Limitations and High-level Approach}.
This paper addresses these limitations by developing a new methodology to bridge the gap between steady-state models and complex grid dynamics. 
In particular, a new model is developed that incorporates the immediate, automated system response into the attacker-defender optimization. 
Notably, our approach decouples the computation of this response from the optimization, this is allowed under certain conditions on the response model. 
We prove that this decoupling preserves optimality, and show how it enables improved scalability and interoperability with arbitrarily complex dynamics models. 
Through this design, our work provides an efficient method for conducting worst-case contingency analysis that captures both steady-state constraints and the immediate effects arising from grid dynamics.

\section{Preliminaries}
\label{sec:preliminaries}

We present the key notations and concepts to model the structure and operation of the electric grid. 
We then describe the formulation of the standard $\Nk$ interdiction model and the primal cut algorithm used for its optimization~\cite{sundar2024exact}.
For convenience, Table~\ref{tab:notation} summarizes the notations used. 

\begin{table}[!t]
\centering
\caption{Electric Grid Notations}
\vspace{-0.3cm}
\label{tab:notation}
\begin{tabular}{ccp{5.5cm}}
\toprule
& Symbol & Definition \\
\midrule
\multirow{6}{*}{\rotatebox{90}{Sets}} 
& $\Graph{} = (\Nodes{}, \Edges{})$     & Graph of the network \\
& $\node{i} \in \Nodes{}$               & Set of buses (nodes) \\
& $\edge{i}{j} \in \Edges{}$            & Set of branches / lines (edges) \\
& $\EdgesFrom{i}$                       & Set of edges directed from bus $\node{i}$ \\
& $\EdgesTo{i}$                         & Set of edges directed to bus $\node{i}$ \\
\midrule
\multirow{6}{*}{\rotatebox{90}{Variables}} 
& $\phaseangle{i}$          & Voltage phase angle at bus $\node{i}$ \\
& $\powergen{i}$            & Active power generated at bus $\node{i}$ \\
& $\powerflow{i}{j}$        & Active power flow on branch $\edge{i}{j}$ \\
& $\powershed{i}$           & Active power shed from bus $\node{i}$ \\
& $\genresponse{i}$         & Generator response at generator $\node{i}$ \\
& $\zload{i}$               & Load shutoff binary variable for loads $\node{i}$ \\
\midrule
\multirow{4}{*}{\rotatebox{90}{Constants}} 
& $\powerload{i}$           & Power consumed at $\node{i}$ \\
& $\maxpowergen{i}$         & Maximum power generated at bus $\node{i}$ \\
& $\invreactance{i}{j}$     & Inverse of the reactance in branch $\edge{i}{j}$ \\
& $\thermalrating{i}{j}$    & Thermal limit of branch $\edge{i}{j}$ \\
\midrule
\multirow{10}{*}{\rotatebox{90}{Interdiction}} 
& $\Attacker{}$ & Attacker Player in the interdiction model \\
& $\Defender{}$ & Defender Player in the interdiction model \\
& $\Response{}$ & Response Model \\
& $\NetworkState{\cdot}$ & Value of variable $(\cdot)$ in network state $\NetworkState{}$ \\
& $\budget$ & Number of lines interdicted by the attacker \\
& $\xline{i}{j}$ & Binary variable for if line $\edge{i}{j}$ is interdicted \\
& $\xlinevec$ & A vector of binary variables $\{\xline{i}{j}\}$ \\
& $\budget[']$ & Number of lines failed in the response model \\
& $\xline[']{i}{j}$ & Binary variable for if line $\edge{i}{j}$ fails in $\Response{}$ \\
& $\xlinevec[']$ & A vector of binary variables $\{\xline[']{i}{j}\}$ \\
\bottomrule
\end{tabular}
\end{table}

\subsection{The Electric Grid Infrastructure}

The electric power grid is a network graph $\Graph{} = (\Nodes{}, \Edges{})$, where nodes are buses, and edges are power lines or branches.
The bus nodes, $\node{i} \in \Nodes{}$, are characterized by their voltage phase angle $\phaseangle{i}$, which is a measure of the shift in voltage relative to a reference bus in the grid. 
Nodes are co-located with generators and load demands that represent power production and consumption. A generator can produce up to $\maxpowergen{i}$ power, while a load demand consumes $\powerload{i}$ power. 
Each branch, $\edge{i}{j} \in \Edges{}$, is defined by its thermal power rating $\thermalrating{i}{j}$, which indicates the maximum power it can safely carry; and a quantity representing the inverse of its reactance, $\invreactance{i}{j}$, which reflects the difficulty of transmitting power.

During operation, the power produced at generators and carried by branches to satisfy load demands determines the \emph{state} of the grid. 
Let $\NetworkState{}$ denote a grid state, consisting of: generator setpoints $\powergen{i}$, indicating the power output of a generator $\node{i}$; and power flows $\powerflow{i}{j}$ indicating the power transmitted across branch $\edge{i}{j}$. 
We define $\NetworkState{\cdot}$ as the value of the variable $(\cdot)$ for a state $\NetworkState{}$. 
Power flows are directed: $\EdgesFrom{i}$ and $\EdgesTo{i}$ denote the sets of edges leaving and entering bus $\node{i}$, with direction assigned arbitrarily.
When load demands cannot be satisfied, e.g., due to inadequate power generation, a grid operator may need to shed load from the system for stability. We use $\powershed{i}$ to denote the amount of power load shed from bus $\node{i}$.

\subsection{The Standard $\Nk$ Interdiction Model}

The $\Nk$ interdiction model analyzes grid vulnerabilities using a two-player game where an \emph{attacker} $\Attacker{}$ aims to fail components to cause the most disruption in the grid, and a \emph{defender} $\Defender{}$ controls the grid's response to the failure. 
This disruption is typically modeled through load shed.
For ease of exposition, this work considers only the failure of transmission lines. Extensions for other grid components, e.g., generators, substations, can be obtained from formulations such as in~\cite{salmeron2009worst} and the $\Nkkp$ formulation described here can be extended to these formulations. 
We next present the standard formulations for the attacker and defender, followed by the primal cut algorithm used to identify worst-case failures.

\textbf{The Standard Attacker Model}.
The standard attacker $\Attacker{}$ in~\eqref{eqn:standard-attacker} identifies the $\budget$ lines whose simultaneous (or near-simultaneous) failure produces the most load shed.
Let $\xline{i}{j}$ denote a binary indicator variable whose value is $1$ if branch $\edge{i}{j}$ has failed, and $0$ otherwise. 
We use $\xlinevec$ to represent the vector consisting of the binary indicator variables $\xline{i}{j}$ for $\edge{i}{j} \in \Edges{}$. 
The attacker's objective in~\eqref{eqn:standard-attacker-objective} is to maximize the load shed response from the defender. 
To do this, the attacker selects $\budget$ lines to disable in the grid, as shown through  Constraints~\eqref{constraint:standard-budget} and~\eqref{constraint:xline-domain}. 
Note that the attacker's choice in $\xlinevec$ is combinatorial, and thus full enumeration is often intractable. 

\makesmallequation*{
\begin{subequations}\label{eqn:standard-attacker}
\begin{align}
\Attacker{\budget} = & \max_{\xlinevec} \quad
\Defender{\xlinevec}
\label{eqn:standard-attacker-objective}
\\
\text{s.t.} \quad
& \sum\nolimits_{\edge{i}{j} \in \Edges{}}
  \xline{i}{j}
  =
  \budget
\label{constraint:standard-budget}
\\
& \xline{i}{j} \in \{0,1\}
&& \forall \edge{i}{j} \in \Edges{}
\label{constraint:xline-domain}
\end{align}
\end{subequations}
}

\textbf{The Standard Defender Model}.
The defender $\Defender{}$ in~\eqref{eqn:standard-defender} controls the grid response to failures, and maximizes the power load that can be safely served to consumers.
The defender's objective is expressed in~\eqref{eqn:standard-defender-objective} as a minimization of the total load shed across all nodes, given the failures $\xlinevec$. 
We then introduce $\xlinehatvec$ as a vector equal to $\xlinevec$ in~\eqref{eqn:xlinevechat}. While this is a redundant constraint, it will be useful for later formulations.
The constraints in~\eqref{eqn:powerflow-domain} and~\eqref{eqn:powergen-domain} are used to capture physical limitations on branches and generators, i.e., power flows on branches cannot violate their thermal ratings, and generators cannot exceed their maximum generation. 
We use~\eqref{eqn:load-shed-domain} to limit the load shed per node. 
The defender is also constrained by physical laws that govern power flows: \emph{Kirchhoff's law} ensures the balance of power flows for each bus in the grid, while \emph{Ohm's Law} relates power flows to voltage phase angles differences for each branch. This is shown in~\eqref{eqn:kirchhoffs-law} and~\eqref{eqn:ohms-law}.
We assume that DC linear approximations are applied for modeling the grid network, similar to the models in~\cite{salmeron2009worst,sundar2021credible}, which approximates AC power with linearized DC power. 

\makesmallequation*{
\begin{subequations}\label{eqn:standard-defender}
\begin{align}
\Defender{\xlinevec{}} ={} & \min \quad
    \sum_{\node{i} \in \Nodes{}}
    \powershed{i}
\label{eqn:standard-defender-objective}
\\
\text{s.t.} \quad
& \xlinehatvec = \xlinevec
\label{eqn:xlinevechat}
\\
& -\thermalrating{i}{j}(1-\xlinehat{i}{j})
  \leq
  \powerflow{i}{j}
  \leq
  \thermalrating{i}{j}(1-\xlinehat{i}{j})
&& \forall \edge{i}{j} \in \Edges{}
\label{eqn:powerflow-domain}
\\
& 0
  \leq
  \powergen{i}
  \leq
  \maxpowergen{i}
&& \forall \node{i} \in \Nodes{}
\label{eqn:powergen-domain}
\\
& 0
  \leq
  \powershed{i}
  \leq
  \powerload{i}
&& \forall \node{i} \in \Nodes{}
\label{eqn:load-shed-domain}
\\
& \powergen{i} {-} \powerload{i} {+} \powershed{i}
  {=}\hspace{-0.2cm}
  \sum_{\edge{i}{j} \in \EdgesFrom{i}} \hspace{-0.3cm} \powerflow{i}{j}
  \hspace{0.05cm} {-}\hspace{-0.2cm}
  \sum_{\edge{j}{i} \in \EdgesTo{i}} \hspace{-0.3cm} \powerflow{j}{i}
&& \forall \node{i} \in \Nodes{}
\label{eqn:kirchhoffs-law}
\\
& \powerflow{i}{j}
  =
  \invreactance{i}{j}
  (\phaseangle{i} - \phaseangle{j})
  (1-\xlinehat{i}{j})
&& \forall \edge{i}{j} \in \Edges{}
\label{eqn:ohms-law}
\end{align}
\end{subequations}
}

\textbf{The Primal Cut Algorithm}.
The attacker $\Attacker{}$ and defender $\Defender{}$ are optimized through the \emph{primal cut algorithm}. 
This algorithm optimizes the attacker and defender in two alternating steps: (i)~the attacker selects $\xlinevec$ representing the $\budget$ lines to fail; and (ii)~the defender re-dispatches generators, re-routes power flows, and sheds loads as necessary to continue serving load demands in the network. 
These steps are repeated iteratively until convergence is reached. 
At the end of each iteration, the total load shed (computed by the defender) is provided back to the attacker. 
The primal cut algorithm accomplishes these steps by constructing \emph{cutting planes} consisting of linear functions, with the goal of defining an upper bound on the total load that was shed by the defender.
We adopt the heuristic version discussed in~\cite{sundar2024exact} to form a cutting plane using coefficients computed for each branch. The approach developed in this paper also applies to the exact method discussed in \cite{sundar2024exact}, albeit with slower convergence to global optimality. 

\SetKw{break}{break}
\begin{algorithm}[!ht]
\small
\KwIn{int $k$, tolerance $\varepsilon$, Grid $\Graph{} = (\Nodes{}, \Edges{})$}
\KwOut{Optimal Attack $\xlinevec[^*]$, Load shed $\powershed{lb}$}
$\Attacker{} \gets \text{InitializeAttackerModel}()$ \\
$\powershed{lb} \gets 0;\,\, \powershed{ub} \gets \infty$; \\
$\xlinevec \gets \text{InitializeAttack}()$ \\
$\xlinevec[^*] \gets \xlinevec$ \\
\While{$\powershed{ub} - \powershed{lb} \geq \varepsilon \powershed{lb}$} {
    $\Defender{} \gets \text{InitializeDefenderModel}()$ \\
    $\powershed{\Defender{}}, \NetworkState[\Defender{}]{} \gets \text{Optimize} \,\,\Defender{\xlinevec}$ \\
    \lIf {$\powershed{lb} < \powershed{\Defender{}}$} {
        $\powershed{lb} \gets \powershed{\Defender{}}{}$;\,\, $\xlinevec[^*] \gets \xlinevec$
    }
    $cut \gets \text{MakeConstraint: } \powershed{ub} \leq \powershed{\Defender{}} +\sum\limits_{\edge{i}{j} \in \Edges{}}{\NetworkState[\Defender{}]{\powerflow{i}{j}}\xline{i}{j}}$ \\
    $\Attacker{}.\text{addConstraint}(cut)$ \\
    $\powershed{ub}, \xlinevec \gets \text{Optimize}\,\, \Attacker{\budget}$ \\
}
\Return $\xlinevec[^*], \powershed{lb}$
\caption{Primal Cut Optimization}
\label{alg:primal-cut-optimization}
\end{algorithm}

Algorithm~\ref{alg:primal-cut-optimization} describes the primal cut optimization algorithm. 
It starts by initializing the attacker model $\Attacker{}$, the lower and upper bounds on load shed, and the initial attack consisting of $\budget$ random lines (lines~1-3). 
The best attack found is denoted by $\xlinevec[^*]$ (line~4). 
The defender model is then executed on $\xlinevec$, and the best lower bound on the load shed is cached. This occurs until convergence, which occurs when the upper and lower bounds on load shed are within a tolerance, $\varepsilon$, of each other (line~5-8).
We then construct the cutting plane, and add it as a constraint to the attacker model (lines~9-10). 
The attacker model is optimized to find an upper bound on the load shed (line~11).
The optimal attack and load shed are returned once the algorithm converges (line~12). 

\section{The $\Nkkp$ Interdiction Problem}
\label{sec:nkkp-interdiction-problem}

One key limitation of the standard $\Nk$ interdiction model is the lack of modeling for the behavior of the grid during and immediately after failures occur. 
That is, the standard model implicitly assumes that the grid transitions directly from its pre-failure state to the computed post-failure steady state.
In practice, however, the grid's immediate response is governed by physical laws and automated local controls, which can cause additional components to fail or trip.
The $\Nkkp$ problem addresses this gap by extending the $\Nk$ framework with a model for the immediate response. This immediate response model accounts for how the initial $\budget$ failures can trigger an additional $\budget[']$ failures in the network. 

\begin{figure}[!ht]
\centering
\includegraphics[width=0.3\textwidth]{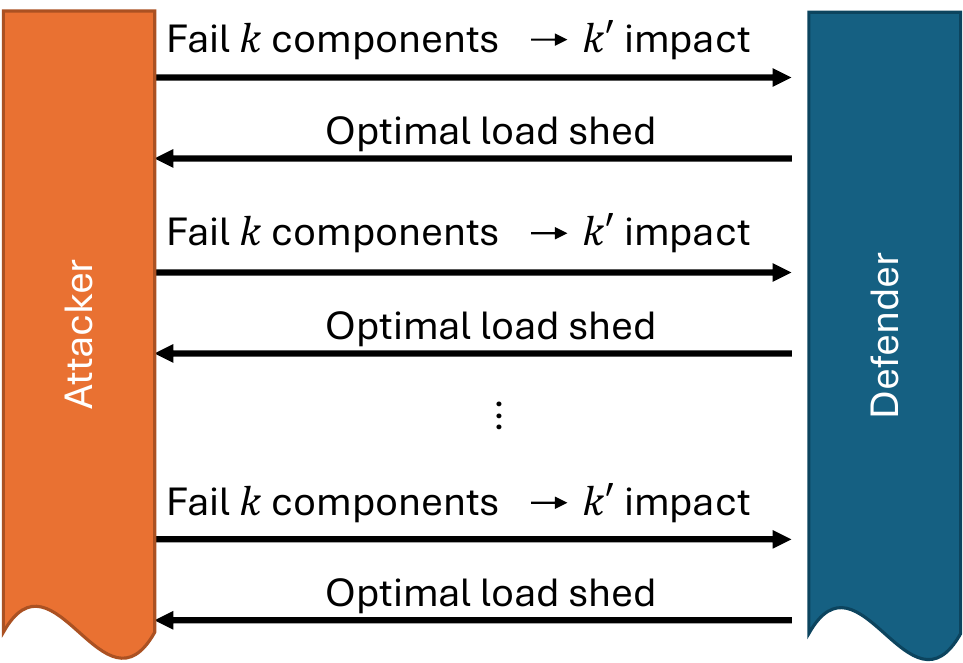}
\vspace{-0.3cm}
\caption{The $\Nkkp$ Interdiction Model}
\label{fig:interdiction-model}
\end{figure}

Next, we define the $\Nkkp$ problem.
Initially, we assume that the grid operates at full capacity, and that all loads in the network are satisfied. 
After some time, $\budget$ transmission lines in the grid simultaneous (or near simultaneously) fail.
As a result, the network state shifts, which causes $\budget[']$ additional lines to subsequently fail, e.g., due to the activation of protection systems. 
The grid operator then applies grid control actions on the joint $\budget \cdot \budget[']$ failures to determine the next steady state for the network. 
In this context, the $\Nkkp$ problem identifies the set of $\budget$ failures that leads to the highest disruption in the grid, measured by the load shed response to the $\budget \cdot \budget[']$ failure. 
We let $\xlinevec$ and $\xlinevec[']$ denote the vectors of binary variables $\xline{i}{j}$ and $\xline[']{i}{j}$, where $\xline{i}{j}$ is $1$ if branch $\edge{i}{j}$ is among the initial $\budget$ failures and $\xline[']{i}{j}$ is $1$ if branch $\edge{i}{j}$ fails during the immediate automated response. 
In Figure~\ref{fig:interdiction-model}, we illustrate the $\Nkkp$ problem setting.
In the following section, we present our disjoint optimization framework that solves this problem by decoupling the response computation from the attacker-defender optimization.

\section{Solution Technique}
\label{sec:approach}

This section details our disjoint approach to solve the $\Nkkp$ problem.
We present this approach through a progression of models, starting with an integrated formulation that directly incorporates the response model into the defender. 
While mathematically correct, this integrated approach introduces significant computational load that makes it difficult to scale to larger grid networks. 
We then show that this computation can be decoupled in the $\Nkkp$ analysis without impacting the underlying optimality and convergence properties of the original problem. 
This decoupling enables an efficient, scalable solution that is also interoperable with complex response. 

\subsection{An integrated model for grid response after failure}
\label{sec:integrated-model}

For illustration, consider an overly simplified model that extends the $\Nk$ analysis to capture the immediate grid response after failures. 
This model examines the impact of re-routed power flows and generator responses if the $\xlinevec$ failures isolate grid components into islands.
Let the defender in this interdiction model be augmented to include a formulation for the additional lines $\xlinevec[']$ (i.e., immediate response failures) that trip as a result of $\xlinevec$. 
For this simple example, we assume that these additional failures occur when power flows on lines exceed their thermal ratings. 
This new defender model is presented in~\eqref{eqn:integrated-defender} below.

\makesmallequation*{
\begin{subequations}\label{eqn:integrated-defender}
\begin{align}
\Defender{\xlinevec{}} ={} & \min \quad
    \sum_{\node{i} \in \Nodes{}} \powershed{i}
\label{eqn:integrated-defender-objective}
\\
\text{s.t.} \quad
& \eqref{eqn:powerflow-domain},
  \eqref{eqn:powergen-domain},
  \eqref{eqn:load-shed-domain},
  \eqref{eqn:kirchhoffs-law},
  \eqref{eqn:ohms-law}
\label{eqn:from-standard-defender}
\\
& \xlinehatvec = \xlinevec + \xlinevec[']
\label{eqn:integrated-xlinevechat}
\\
& \powergen[\tilde{p}]{i} = \powergen[\phi]{i} + \genresponse{i}
&& \forall \node{i} \in \Nodes{}
\label{eqn:generator-response-value}
\\
& \genresponse{i}(1-\xline{i}{j})
  =
  \genresponse{j}(1-\xline{i}{j})
&& \forall \edge{i}{j} \in \Edges{}
\label{eqn:equal-generator-response}
\\
& \xline[']{i}{j}
  \leq
  |\powerflow[\tilde{p}]{i}{j}| / \thermalrating{i}{j}
&& \forall \edge{i}{j} \in \Edges{}
\label{eqn:force-zero-if-pf_lt_rate_a}
\\
& \largeM \xline[']{i}{j}
  \geq
  |\powerflow[\tilde{p}]{i}{j}| - \thermalrating{i}{j}
&& \forall \edge{i}{j} \in \Edges{}
\label{eqn:force-one-if-pf_gt_rate_a}
\\
& \xline{i}{j} + \xline[']{i}{j}
  \leq
  1
&& \forall \edge{i}{j} \in \Edges{}
\label{eqn:sum-xline-one}
\\
& 0
  \leq
  \xline[']{i}{j}
  \leq
  1
&& \forall \edge{i}{j} \in \Edges{}
\label{eqn:xlinep-domain}
\\
& \powergen[\tilde{p}]{i} {-} \powerload{i}
  =\hspace{-0.2cm}
  \sum_{\edge{i}{j} \in \EdgesFrom{i}} \hspace{-0.2cm} \powerflow[\tilde{p}]{i}{j}
  {-}\hspace{-0.2cm}
  \sum_{\edge{j}{i} \in \EdgesTo{i}} \hspace{-0.2cm} \powerflow[\tilde{p}]{j}{i}
&& \forall \node{i} \in \Nodes{}
\label{eqn:kirchhoffs-law-repsonse}
\\
& \powerflow[\tilde{p}]{i}{j}
  =
  \invreactance{i}{j} (\phaseangle{i} - \phaseangle{j})(1-\xline{i}{j})
&& \forall \edge{i}{j} \in \Edges{}
\label{eqn:ohms-law-response}
\end{align}
\end{subequations}
}

The new defender model keeps the standard defender's objective and most of its constraints in~\eqref{eqn:integrated-defender-objective} and~\eqref{eqn:from-standard-defender}. 
One key difference is that the failures $\xlinehatvec$ are now set to $\xlinevec + \xlinevec[']$ in~\eqref{eqn:integrated-xlinevechat}.
In doing so, the defender will take control actions to minimize the disruption with respect to the combined failure. 

Physical laws in the grid dictate that the system stays in balance, with generators instantaneously adjusting their production in response to imbalances -- in this case, when $\xlinevec$ separates the network into two or more pieces (islands). This is sometimes referred to as droop or frequency response. 
If an island experiences a significant mismatch between generation and load, the resulting frequency deviations can trigger protective systems and cause other component failures beyond those in $\xlinevec$.
To model droop, we introduce a generator response variable $\genresponse{i}$, and assign the next setpoint for each generator in~\eqref{eqn:generator-response-value}. 
Here, $\powergen[\tilde{p}]{i}$ is a generator setpoint variable used during the immediate automated response, and $\powergen[\phi]{i}$ is a constant denoting the generator setpoint prior to $\xlinevec$ failing.
We enforce a uniform droop response within each island formed after the initial failures $\xlinevec$ by requiring that $\genresponse{i} = \genresponse{j}$ for edges $\edge{i}{j}$ that have not failed, in~\eqref{eqn:equal-generator-response}. While this is an oversimplified model of droop control, we note that even this model introduces computational complexity to the defender model that is only exacerbated for more sophistical models of droop response.

Finally, as mentioned earlier, power flows on branches that have not failed may potentially exceed their thermal rating and trip. 
Similar to above, we introduce $\powerflow[\tilde{p}]{i}{j}$ as the power flow variable during the immediate automated response. 
To model the tripping of overloaded lines, we use~\eqref{eqn:force-zero-if-pf_lt_rate_a} and~\eqref{eqn:force-one-if-pf_gt_rate_a} to assign $\xline[']{i}{j}$ based on whether the power flow magnitude exceeds the thermal rating.
When $|\powerflow[\tilde{p}]{i}{j}| \leq \thermalrating{i}{j}$, the constraint in~\eqref{eqn:force-zero-if-pf_lt_rate_a} requires $\xline[']{i}{j}$ to be $0$. Conversely, when $|\powerflow[\tilde{p}]{i}{j}| > \thermalrating{i}{j}$, the constraint in~\eqref{eqn:force-one-if-pf_gt_rate_a} requires $\xline[']{i}{j}$ to be $1$. 
In each case, only one constraint actively restricts $\xline[']{i}{j}$; the other becomes redundant and is automatically satisfied.\footnote{The absolute value in these constraints can be linearized using binary variables for when power flow is positive, but is omitted for conciseness.} We use~\eqref{eqn:sum-xline-one} and~\eqref{eqn:xlinep-domain} to define bounds for $\xline[']{i}{j}$ and ensure that at most one of $\xline{i}{j}$ and $\xline[']{i}{j}$ equals $1$ across all branches.
Finally, during the immediate automated response, power flows and generation in the grid still follows laws of physics; however, load cannot be shed at this timescale. These laws are expressed in~\eqref{eqn:kirchhoffs-law-repsonse} and~\eqref{eqn:ohms-law-response}.

\subsection{Decoupling the response computation from steady-state optimization}

While the integrated model in~\eqref{eqn:integrated-defender} captures the immediate automated response following the initial failures $\xlinevec$, it greatly increases the computational complexity of the defender by coupling $\xlinevec$ and $\xlinevec[']$. 
As a result, when more realistic and/or complex response models are used, e.g., those that use more detailed models of droop or details of grid dynamics (modeled with partial differential equations (PDEs)), they can introduce additional non-convexities to the defender optimization model. 
This can compromise the computational efficiency, convergence, and optimality guarantees of the primal cut algorithm.

However, we observe the following key insight, that the primal cuts generated in Algorithm~\ref{alg:primal-cut-optimization} are constructed entirely in the space of $\xlinevec$ and based on the formulation of the standard defender in~\eqref{eqn:standard-defender}. 
This allows the computation of $\xlinevec[']$ to be pulled out of the defender problem. 
Hence, the immediate automated system response can be captured by computing $\xlinevec[']$ separately, and passing the combined failures $\xlinevec + \xlinevec[']$ into the standard defender model $\Defender{}$. 
This yields the construction of a defender model that operates on the combined failures $\xlinevec + \xlinevec[']$ and produces the \emph{same cut coefficients for the primal cut algorithm}. 

We next present a new attacker and response model that implements this insight in~\eqref{eqn:decoupled-attacker}. 
This formulation will be integrated with Algorithm~\ref{alg:primal-cut-optimization}, where the response model $\Response{}$ is invoked during each iteration to compute $\xlinevec[']$ before evaluating the defender.
That is, the attacker continues to select exactly $\budget$ initial lines, and the defender is now evaluated on the combined failures $\xlinevec+\xlinevec[']$, as expressed in~\eqref{eqn:decoupled-attacker-objective}. 
Note that these combined failures are passed to the defender, which models how $\xlinehatvec$ is set in~\eqref{eqn:integrated-xlinevechat}.
The response model $\Response{}$ computes the additional failures $\xlinevec[']$ in~\eqref{eqn:response-model-xlinevec}.

\makesmallequation*{
\begin{subequations}\label{eqn:decoupled-attacker}
\begin{align}
\Attacker{\budget} ={} & \max_{\xlinevec} \quad
    \Defender{\xlinevec + \xlinevec[']}
\label{eqn:decoupled-attacker-objective}
\\
\text{s.t.} \quad
& \sum_{\edge{i}{j} \in \Edges{}}
  \xline{i}{j}
  =
  \budget
\label{constraint:decoupled-budget}
\\
& \xlinevec['] = \Response{\xlinevec}
\label{eqn:response-model-xlinevec}
\end{align}
\end{subequations}
}

One implementation for the response model $\Response{}$ that reformulates the integrated model is provided in Algorithm~\ref{alg:decoupled-response-model}. 
Given the grid $\Graph{}$ and a set of failures $\xlinevec$, we start by constructing $\Graph[*]{}$, the post-failure system consisting of nodes $\Nodes{}$ and edges that have not failed (line~1). 
As mentioned earlier, this can potentially create islands in the grid. 
$\xlinevec[']$ is also initialized to the empty set (line~2). 
For each island $\Graph[isl]{}$ in the graph $\Graph[*]{}$, we explicitly compute $\genresponse{}$, its constant generator response. This algorithmically models the constraints stated in~\eqref{eqn:generator-response-value} and~\eqref{eqn:equal-generator-response}.
The new generator setpoints for the grid network are then computed as a function of the previous network state (lines~5-7).
The response model then solves for the power flow on lines (line~8), which internally must satisfy the constraints stated in~\eqref{eqn:kirchhoffs-law-repsonse} and~\eqref{eqn:ohms-law-response}.
To determine the $\budget[']$ failures that are in $\xlinevec[']$, we filter the branches in the network whose power flows $\powerflow{i}{j}$ exceed their thermal rating limits $\thermalrating{i}{j}$ (lines~9-10). This captures the constraints stated in~\eqref{eqn:force-zero-if-pf_lt_rate_a} and~\eqref{eqn:force-one-if-pf_gt_rate_a}.
The algorithm lastly returns $\xlinevec[']$ as the set of additional failures (line~11).

\makesquishalgorithm[18pt]{
\begin{algorithm}[!t]
\small
\KwIn{Graph $\Graph{}{} = (\Nodes{}{}, \Edges{}{})$, Previous State $\NetworkState[prev]{}$, Lines $\xlinevec$}
\KwOut{Response Model impact $\xlinevec{'}$}
$\Graph[*]{} \gets DisableComponents(\xlinevec, \Graph{}{})$\\
$\xlinevec{'} \gets \emptyset$ \\
\tcp{Go through each island in $\Graph[*]{}$}
\For{$\Graph[isl]{} \gets ComputeIslands(\Graph[*]{})$} {
    \If{$\Nodes[isl]{} \neq \emptyset$} {
        \tcp{Find new setpoints for generators from $\NetworkState[prev]{}$}
        $\genresponse{} \gets \sum_{\node \in \Nodes[isl]{}}(\powerload{i} - \powergen{i}) ~/~ \left\vert \Nodes[isl]{} \right\vert $ \\
        $new\_gen\_setpoints \gets map()$ \\
        \lFor{$\node{i} \in \Nodes{}$} {
            $new\_gen\_setpoints[\node{i}] \gets \NetworkState[prev]{\powergen{i}} + \genresponse{i}$
        }
    
        \tcp{Determine failures within component}
        $\NetworkState{} \gets solve\_pf(\Graph[isl]{}, generator\_setpoints=new\_gen\_setpoints)$ \\
        \For{$\edge{i}{j} \gets \Edges[isl]{}$} {
            \lIf{$\left| \NetworkState{\powerflow{i}{j}} \right| > \thermalrating{i}{j}$} {
                $\xlinevec['] \gets \xlinevec['] \cup \{\edge{i}{j}\}$
            }
        }
    }
}
\Return $\xlinevec{'}$\\
\caption{Decoupled Response Model}
\label{alg:decoupled-response-model}
\end{algorithm}
}

Note that while Algorithm~\ref{alg:decoupled-response-model} implements the response logic from our integrated model, the decoupled framework allows for much more general response models.
In particular, this decoupled response model operates as a black-box for computing $\xlinevec[']$ while preserving the original guarantees of the algorithm. 
As long as $\Response{}$ produces a mapping from $\xlinevec$ to $\xlinevec[']$ and operates independently of the defender's steady-state optimization, \emph{any arbitrarily complex response model can be employed} by replacing lines 1-10 with an arbitrary response model, such as a transient simulation. This leads to the full $\Nkkp$ model as formalized in Algorithm~\ref{alg:full-decoupled}.

\makesquishalgorithm[20pt][4pt]{
\begin{algorithm}
\small
\KwIn{Graph $\Graph{}{} = (\Nodes{}{}, \Edges{}{})$, Previous State $\NetworkState[prev]{}$, int $\budget$}
\KwOut{Worst-case failure event $(\powershed{}, \xlinevec + \xlinevec{'})$}
$\xlinevec \gets \Attacker{\budget}$ \\
$\xlinevec{'} \gets \Response{\xlinevec}$ \\
$\powershed{} \gets \Defender{\xlinevec + \xlinevec{'}}$ \\
\Return $(\powershed{}, \xlinevec + \xlinevec{'})$ \\
\caption{Evaluating the $\Nkkp$ Model}
\label{alg:full-decoupled}
\end{algorithm}
}

\subsection{Theoretical Evaluation}

We next show that the decoupling above does not alter the underlying optimization model.
We present this through the lemma below, and then state several remarks.

\textbf{Lemma}.
\emph{
Assume that the response model $\Response{}$ computes $\xlinevec{'}$ as a function of only $\xlinevec$, and that the defender model $\Defender{\xlinevec}$ is deterministic. 
Then, for any given $\xlinevec$, if $\Cut$ denotes the cut coefficients generated by a deterministic optimization model, the same cut can be subsequently generated when that optimization is treated as a black box.  
}

\textbf{Proof}.
Let $\Defender[1]{}$ and $\Defender[2]{}$ denote two constructions of the defender model considering the response model $\Response{}$.
The first defender model $\Defender[1]{}$ is integrated with $\Response{}$; it is given $\xlinevec$ as input, and must compute both $\xlinevec[']$ and the next grid state.
The second defender model $\Defender[2]{}$ is decoupled from $\Response{}$; it is given the combined failure $\xlinevec + \xlinevec[']$ as input, and then determines the next grid state. 
Let $\Cut[1]$ and $\Cut[2]$ denote the cuts that are generated by each defender model $\Defender[1]{}$ and $\Defender[2]{}$, respectively.
Since $\Response{}$ computes $\xlinevec[']$ from $\xlinevec$ deterministically, the value of $\xlinevec[']$ in $\Defender[1]{}$ is fixed and enters the model only through the combined failure $\xlinevec + \xlinevec[']$.
This input is what is provided to $\Defender[2]{}$.
Thus, it follows that the two constructions share the same parameters, the same feasible set, and the same objective, and therefore represent the same defender problem.

Let $S^{\star}$ denote the set of grid states that minimize the total load shed for this problem, which is common to both defender models.
As discussed in~\cite{sundar2024exact}, the coefficients of the cuts $\Cut[1]$ and $\Cut[2]$ are determined by $\NetworkState{\powerflow{i}{j}}$, the power flows on each branch $\edge{i}{j}$ in the state of the grid computed by the defender.
Let $\NetworkState[1]{}$ denote the grid state computed by $\Defender[1]{}$, so that the coefficients of $\Cut[1]$ are those induced by the power flows in $\NetworkState[1]{}$.
By definition, the objective of the defender is to minimize the load shed in the grid network, and since $\Defender[1]{}$ is solved to global optimality, $\NetworkState[1]{}$ belongs to $S^{\star}$.
The state $\NetworkState[1]{}$ is then an optimal solution of the problem solved by $\Defender[2]{}$ as well.
Consequently, $\Defender[2]{}$, when solved to global optimality is also able to return $\NetworkState[1]{}$, and the cut it generates from that state has the same coefficients as $\Cut[1]$.
Therefore, the cut coefficients generated by a deterministic optimization model can be produced even when the model is treated as a black box.
\qed

There are two major consequences of this lemma. 
First, we note that when the response model is unique and deterministic, then the two methods will produce the same solution. 
This is because the two constructions will match at each iteration of the cutting plane algorithm. 
Second, when the response model can yield a non-unique solution, e.g., due to internal stochasticity in the optimizer or solution degeneracy, the two approaches are equivalent at the formulation level.
This behavior is common across any two solvers that are applied to a problem with non-unique solutions, e.g., a MIP solver on a problem formulation with multiple optimal solution. 
\section{Experiments}
\label{sec:experiments}

We next present the evaluation of the decoupled approach on several benchmark power grid networks of varying scales. 
Details of each network and our experimental setup is provided in Section~\ref{sec:experimental-setup}.
Our experiments first measure the performance of the decoupled approach and validate the impact of capturing the immediate response for the interdiction analysis (Section~\ref{sec:decoupled-approach-performance}). 
We then study the selection of components failed across the $\Nkkp$ models, and how it affects the operating margin left on the surviving network (Section~\ref{sec:selection-failed-components}).
Lastly, we explore the interoperability of our model with other more complex response models (Section~\ref{sec:decoupled-approach-interoperability}). 

\begin{table}[!ht]
\caption{Electric Power Grid Case Studies}
\vspace{-0.3cm}
\label{tab:grid-case-studies}
\begin{tabular}{lcccc}
\toprule
Case Study & \#Buses & \#Lines & Total & Total API \\
 & & & Load (MW) & Load (MW) \\
\midrule
case14\_ieee & 14 & 20 & 259.0 & 463.0 \\
case24\_ieee\_rts & 24 & 38 & 2850.0 & 5470.4 \\
case30\_ieee & 30 & 41 & 283.4 & 471.2 \\
case39\_epri & 39 & 46 & 6254.2 & 10093.5 \\
case73\_ieee\_rts & 73 & 120 & 8550.0 & 16416.4 \\
case89\_pegase & 89 & 210 & 5727.9 & 6786.8 \\
case118\_ieee & 118 & 186 & 4242.0 & 6874.8 \\
case162\_ieee\_dtc & 162 & 284 & 7239.1 & 7647.8 \\
case179\_goc & 179 & 263 & 30326.6 & 71840.9 \\
case197\_snem & 197 & 286 & 1474.1 & 2620.8 \\
case200\_activ & 200 & 245 & 1475.7 & 2456.4 \\
case240\_pserc & 240 & 448 & 144179.7 & 180688.5 \\
case300\_ieee & 300 & 411 & 23525.9 & 26426.7 \\
case500\_goc & 500 & 728 & 17772.9 & 27536.9 \\
case588\_sdet & 588 & 686 & 10661.1 & 14157.3 \\
case793\_goc & 793 & 913 & 13198.3 & 18542.1 \\
case3022\_goc & 3022 & 4135 & 57997.5 & 68545.7 \\
case4661\_sdet & 4661 & 5997 & 88203.6 & 104714.0 \\
\bottomrule
\end{tabular}
\end{table}

\subsection{Experimental Setup}
\label{sec:experimental-setup}

Our evaluation considers several benchmark power grid case studies from the PGLib-OPF v23.07 library~\cite{babaeinejadsarookolaee2019power}.
Table~\ref{tab:grid-case-studies} presents basic information about each network, operated under typical conditions (the base case benchmarks as originally specified), and under heavy load (API cases). 
All networks report power in the MW base. 
We initialize each network to the steady state determined by the DC optimal power flow (DC-OPF) for the system.
Our experiments are conducted on a M1 Macbook Pro with 16 GB of memory and 10 CPU cores. The code was implemented in Julia using JuMP~\cite{Lubin2023} and PowerModels~\cite{powermodels}. We also used the Gurobi optimization solver v12.0.2~\cite{gurobi}, with a $1\%$ optimality tolerance and $30$ minute timeout. 
The implementations of each model are published on GitHub~\cite{temporal_nk-github}. 

\subsection{Performance of the decoupled attacker-response-defender model}
\label{sec:decoupled-approach-performance}

We first explore the performance of the decoupled attacker-response-defender model by measuring its load shed response and computation time across an increasing number of failures. 

\begin{figure*}[b]
  \centering
  \subfloat[Standard $\Nk$ \label{fig:exp5-standard-reg}]{%
    \includegraphics[width=0.333\textwidth]{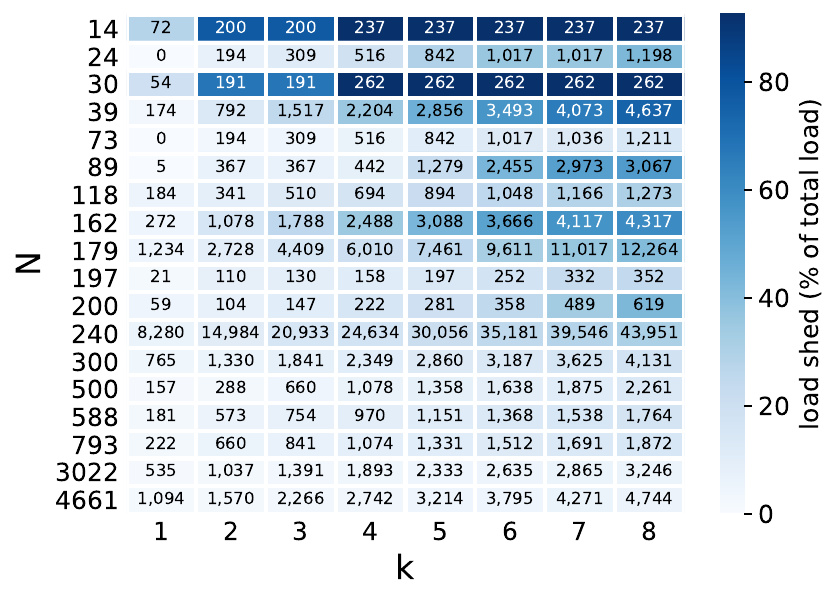}}\hfill
  \subfloat[Decoupled $\Nkkp$ \label{fig:exp5-temporal-kkp-reg}]{%
    \includegraphics[width=0.333\textwidth]{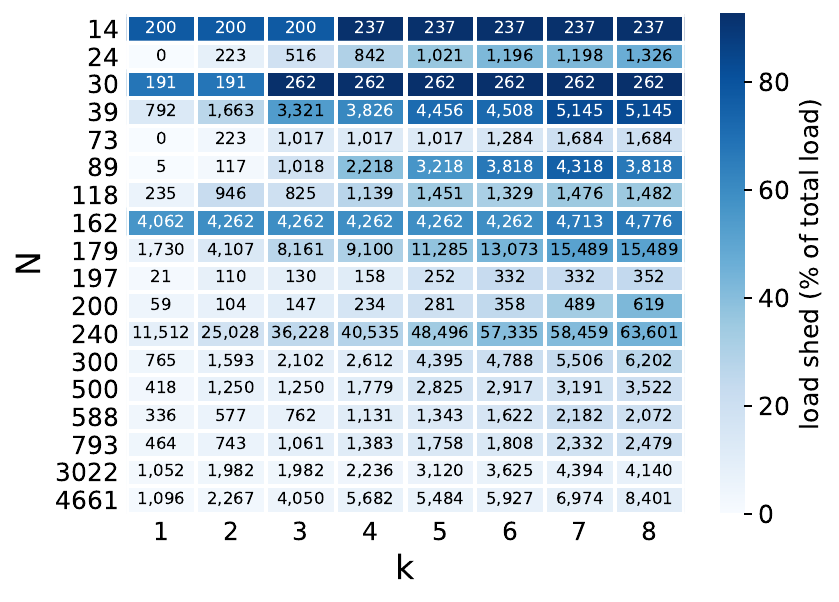}}\hfill
  \subfloat[Integrated $\Nkkp$ \label{fig:exp5-temporal-kkp-joint-reg}]{%
    \includegraphics[width=0.333\textwidth]{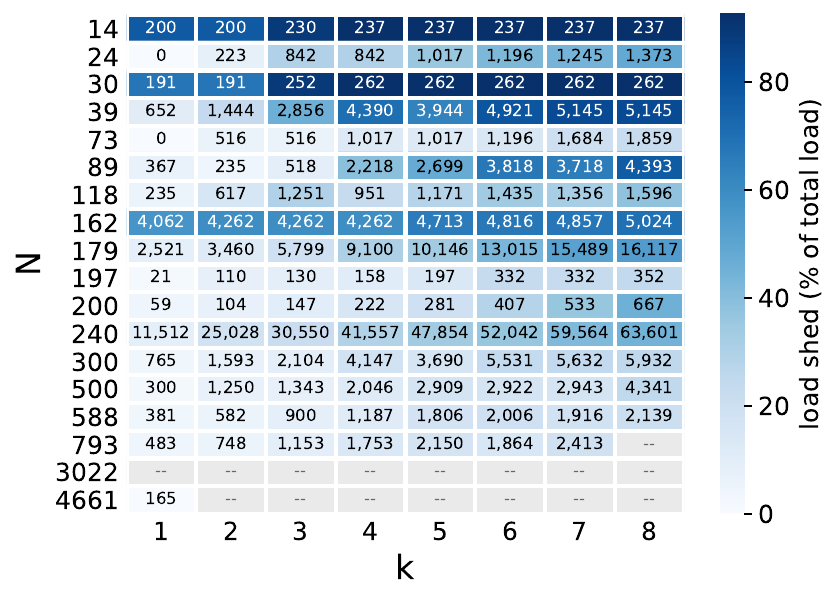}}\\
  \subfloat[Standard $\Nk$ (API)\label{fig:exp5-standard-api}]{%
    \includegraphics[width=0.333\textwidth]{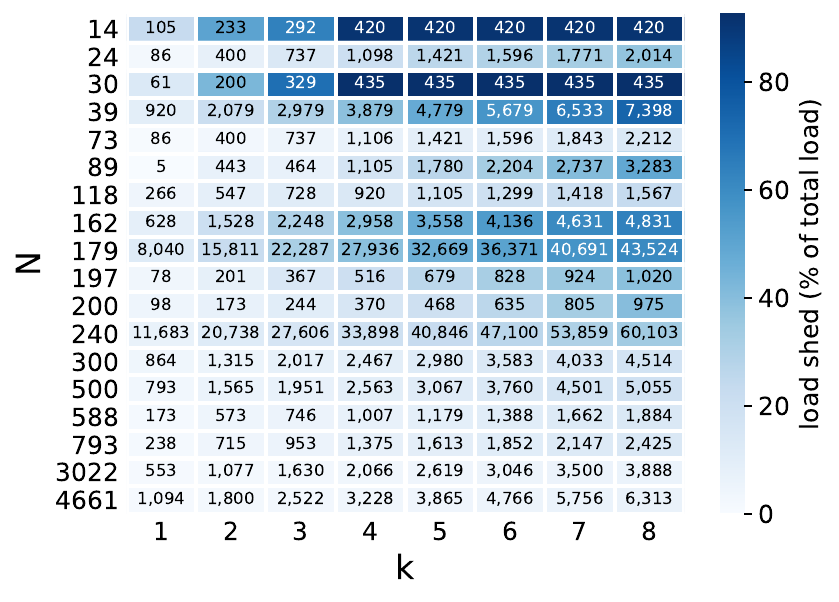}}\hfill
  \subfloat[Decoupled $\Nkkp$ (API)\label{fig:exp5-temporal-kkp-api}]{%
    \includegraphics[width=0.333\textwidth]{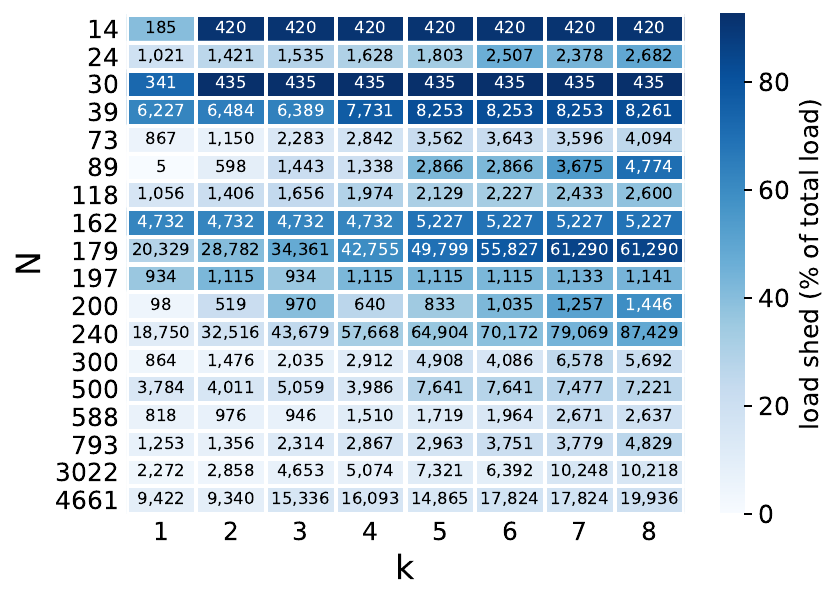}}\hfill
  \subfloat[Integrated $\Nkkp$ (API)\label{fig:exp5-temporal-kkp-joint-api}]{%
    \includegraphics[width=0.333\textwidth]{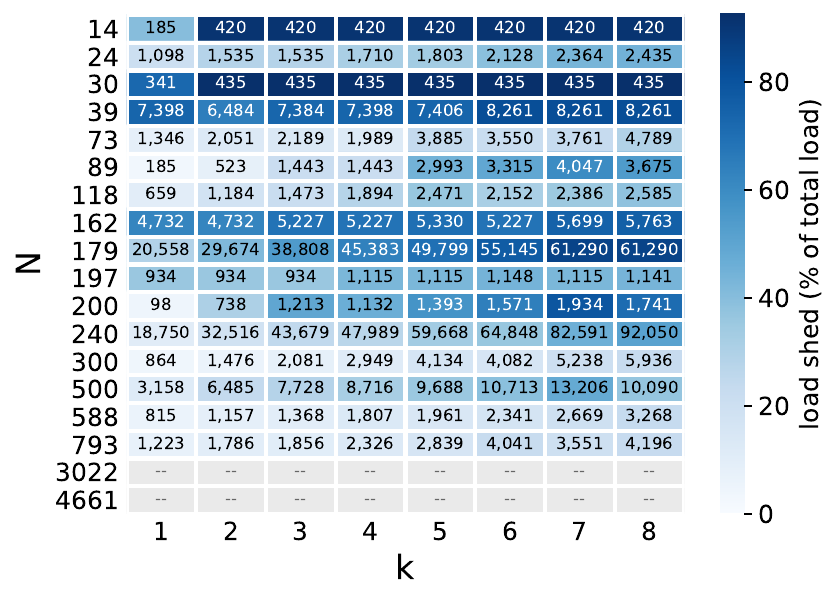}}
  \caption{Load shed (MW) across the standard $\Nk$ model, and the decoupled/integrated $\Nkkp$ models for multiple networks of varying scale and on number of failures. Darker colors indicate higher load shed for the system. Both $\Nkkp$ models identify failures that cause larger load shed than the standard $\Nk$ model, and the two agree closely with each other. (a)-(c) show results for regular cases, and (d)-(f) show results for the API cases. Gray cells mark runs that did not finish within the time limit.}
  \label{fig:load-shed-comparison}
\end{figure*}

\subsubsection{Load shed response}

Figures~\ref{fig:load-shed-comparison} shows heatmaps of the load shed response for failures ranging from $\budget {\,=\,} 1$ to $8$. 
For ease of comparison, color is used to indicate the proportion of load shed response for combinations of network and failure size $\budget$. 
We observe that the load shed typically increases with $\budget$, and that the $\Nkkp$ interdiction models find a disruption at least as large as the standard $\Nk$ model.  
Under typical operating conditions, the load shed identified by the decoupled $\Nkkp$ model was on average $60.8\%$ larger than the standard $\Nk$ model, while for the API cases, it was an average of $135.6\%$ larger. 
This gap is widest at small values of $\budget$, where the standard model has fewer choices of failures: at $\budget {=} 1$ the mean increase is $161.5\%$ for the normal operation cases and $423.3\%$ for the API cases, and it narrows monotonically to $24.6\%$ and $54.6\%$ at $\budget {=} 8$.
This difference comes as a result of the lines failed during the initial choice of failures ($\budget$) and its impact $(\budget['])$, which is explored further in Sec.~\ref{sec:selection-failed-components}.
We additionally observe that the integrated and decoupled $\Nkkp$ models generally produce the same solution on most instances, with discrepancies arising when branch flows are within solver tolerances of their thermal limits. Since this threshold determines whether a line trips during the response model, the induced set $\xlinevec[']$ can differ slightly, affecting the total load shed.
While these effects can be mitigated by tightening tolerances, they will differ from one case study to another; the selection of the correct floating point tolerance is out of the scope of this paper.
Together, this motivates the need to study the expanded $\Nkkp$ model for grid resilience.

\subsubsection{Computation time}

\begin{figure}[!t]
  \centering
  \includegraphics[width=0.36\textwidth]{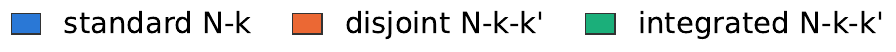}\\
  \subfloat[PGLib-OPF Cases\label{fig:exp6-runtimes-reg}]{%
    \includegraphics[width=0.40\textwidth]{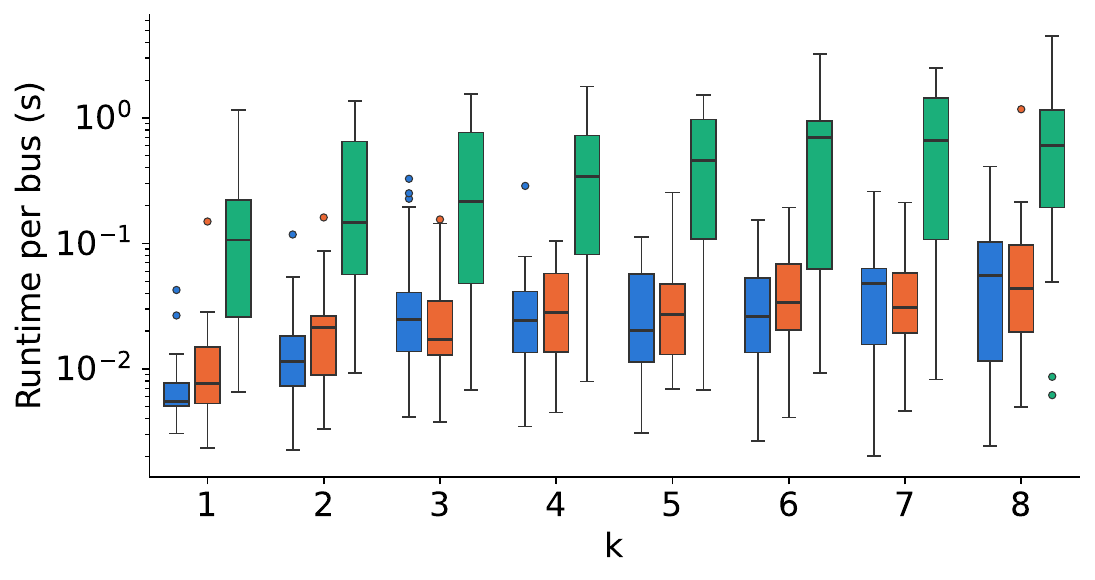}}\\
  \subfloat[PGLib-OPF API Cases\label{fig:exp6-runtimes-api}]{%
    \includegraphics[width=0.40\textwidth]{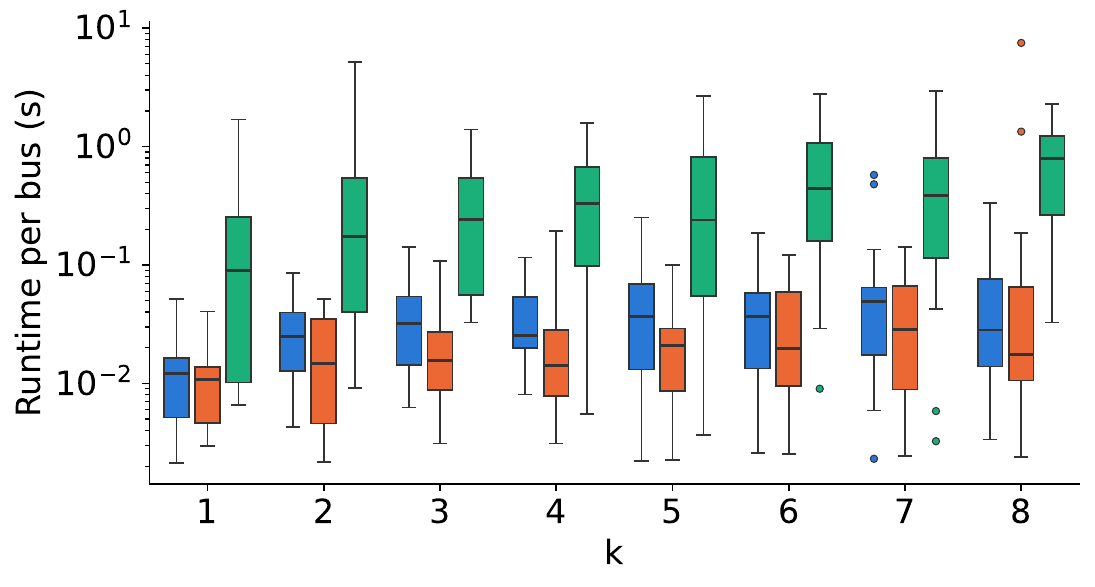}}
  \caption{Box plot on the computational runtime, normalized by bus count, as a function of $\budget$ across all cases. At each $\budget$, the three boxes are the spread of the normalized runtime. This shows how the decoupled $\Nkkp$ model is more computationally efficient than the integrated $\Nkkp$ model.}
  \label{fig:runtime-comparison}
\end{figure}

We next measure the end-to-end computation time taken for one complete run to identify worst-case failures in each network. 
Figure~\ref{fig:runtime-comparison} shows a boxplot of the average time taken by each model across an increasing number of failures from $1$ to $8$. The reported numbers are normalized by the number of bus nodes in the network.
In general, as the value of $\budget$ increases, each model takes longer to execute. 
For the regular cases, the mean normalized runtime grows from $9.3$, $18.1$, and $210.8$ ms at $\budget {\,=\,} 1$ to $92.6$, $122.8$, and $929.7$ ms at $\budget {\,=\,} 8$ for the standard, decoupled, and integrated models, respectively. 
This indicates that the cost of the standard and decoupled models are comparable, while the integrated model is significantly more expensive; this causes the integrated model to begin timing out once the network size and number of failures grows. In particular, our empirical results show that the integrated model fails to solve cases larger than 300 buses within the time limit. 
The results translate to the API cases in a similar manner.
Overall, this shows the efficiency and scalability of the proposed decoupled approach.

\subsection{Selection of Failed Components across the $\Nkkp$ Models}
\label{sec:selection-failed-components}

\begin{figure}
  \centering
  \subfloat[\label{fig:kshared}]{%
    \includegraphics[width=0.4\textwidth]{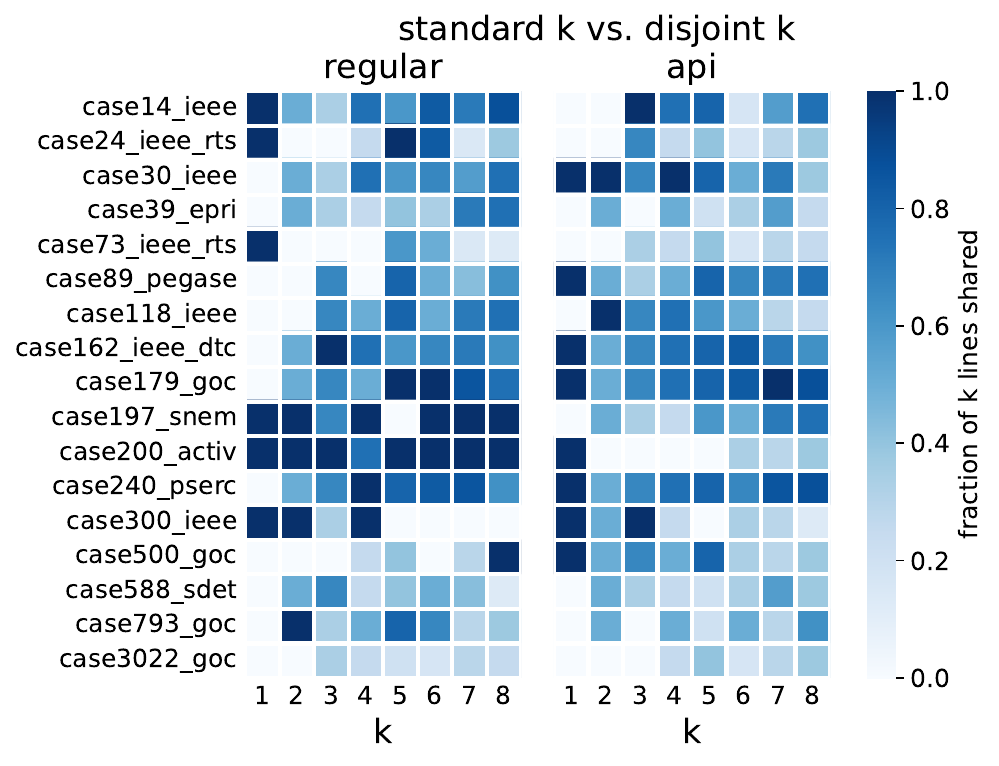}}\hfill
  \subfloat[\label{fig:kcontain}]{%
    \includegraphics[width=0.4\textwidth]{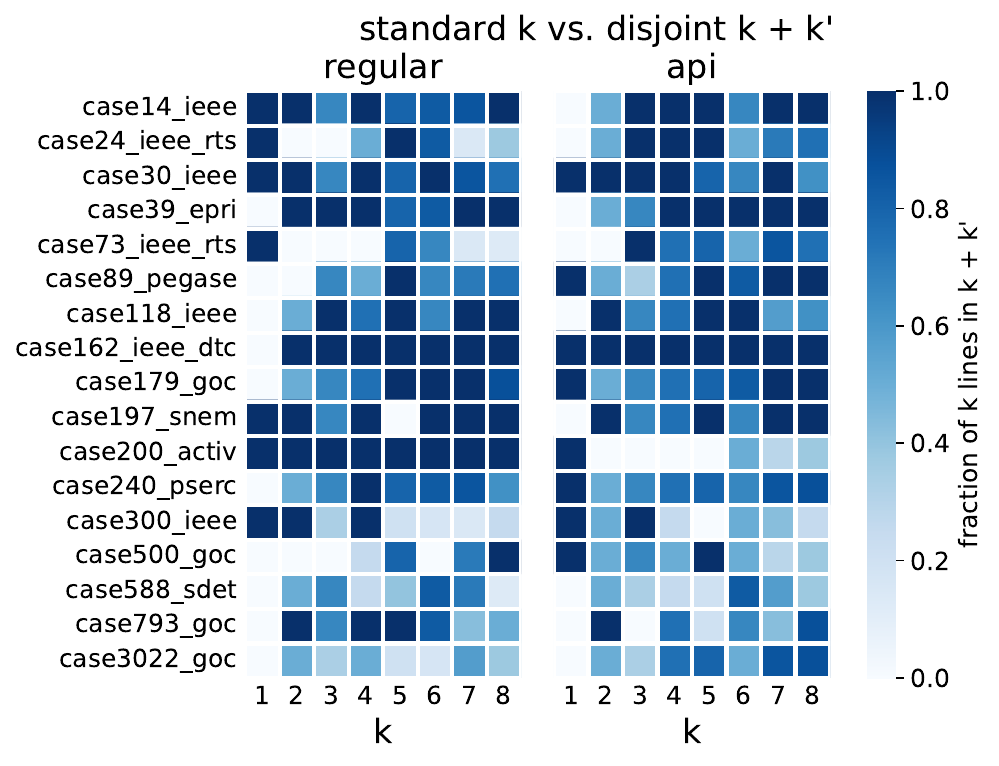}}
  \caption{Overlap between the failures selected by the standard $\Nk$ and decoupled $\Nkkp$ model. (a) Fraction of the standard model's $\budget$ failures also chosen among the decoupled model's $\budget$ failures; (b) Fraction contained in its combined $\budget+\budget[']$ failures. Darker colors indicate greater overlap.}
  \label{fig:klines-comparison}
\end{figure}

Our next experiment explores how the choice in the $\budget$ lines failed differs between the standard $\Nk$ model and the decoupled $\Nkkp$ model.
Figure~\ref{fig:kshared} depicts a heatmap where cells are colored based on the proportion of the $\budget$ failures selected in the standard $\Nk$ model that are also selected in the $\Nkkp$ model. 
Figure~\ref{fig:kcontain} then presents the fraction of lines selected in the standard $\Nk$ model that lie in either the $\budget$ or $\budget[']$ failures in the $\Nkkp$ model.
We observe that the choice of the initial $\budget$ failures differs across the standard $\Nk$ model and decoupled $\Nkkp$ models, for multiple sizes of network and number of failures.
On average, only $49.6\%$ of the standard model's $\budget$ targets are also selected by the decoupled model on the regular cases, and $45.5\%$ on the API cases; the two models select entirely disjoint target sets in $22\%$ and $17\%$ of instances, and agree completely in only $19\%$ and $10\%$.
However, after considering the induced failures, this gap becomes larger: only $63.2\%$ (regular) and $64.1\%$ (API) of the standard model's targets fall somewhere in the decoupled model's combined $\budget + \budget[']$ failure, which is fully containing in $38\%$ and $33\%$ of instances.
Together, these results demonstrate that considering the $\budget[']$ failures changes the initial failure, and help to further motivate the need for the $\Nkkp$ model. 
In some cases, the failure selected by the standard $\Nk$ model is part of the $\budget[']$ failures identified in the $\Nkkp$ model, while in others, a completely different, larger disruption was found by considering $\budget+\budget[']$.

\begin{figure}[!t]
\centering
\includegraphics[width=0.485\textwidth]{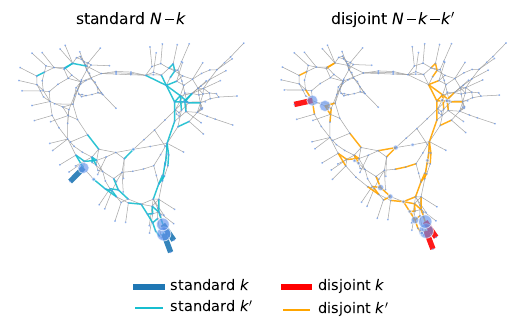}
\vspace{-0.3cm}
\caption{The $\budget {=} 3$ failure on Case 240 PSERC. The left graph shows failures from the standard $\Nk$ model: $\budget$ targets are in blue while the $\budget[']$ failures it does not account for are in cyan. The right graph shows failures from the decoupled $\Nkkp$ model: $\budget$ failures are in red, and induced $\budget[']$ failures are in orange. Circles at nodes represents the amount of load shed. }
\label{fig:kp-example}
\end{figure}

Figure~\ref{fig:kp-example} shows an illustrative example of this difference for the Case 240 PSERC benchmark grid at $\budget {=} 3$, the smallest budget at which the two models select different failures.
In particular, both models share two failed lines, but differ on the third: the selection by the standard $\Nk$ model sheds
sheds $20{,}933$ MW ($14.5\%$ of system load), while the selection by the $\Nkkp$ model sheds 
$36{,}228$ MW ($25.1\%$ of system load). 
This load shed is visualized as larger circles in the figure. 

\subsection{Examining Network Congestion after Failures}
\label{sec:network-congestion}

\begin{figure}
  \centering
  \includegraphics[width=0.36\textwidth]{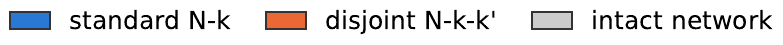}\\
  \subfloat[PGLib-OPF Cases\label{fig:congestion-reg}]{%
    \includegraphics[width=0.40\textwidth]{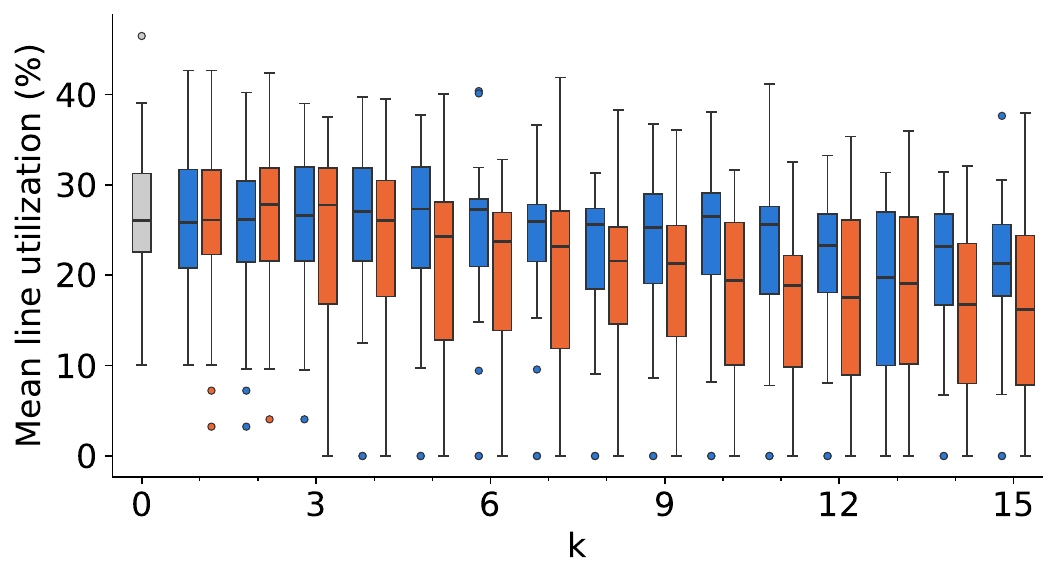}}\\
  \subfloat[PGLib-OPF API Cases\label{fig:congestion-api}]{%
    \includegraphics[width=0.40\textwidth]{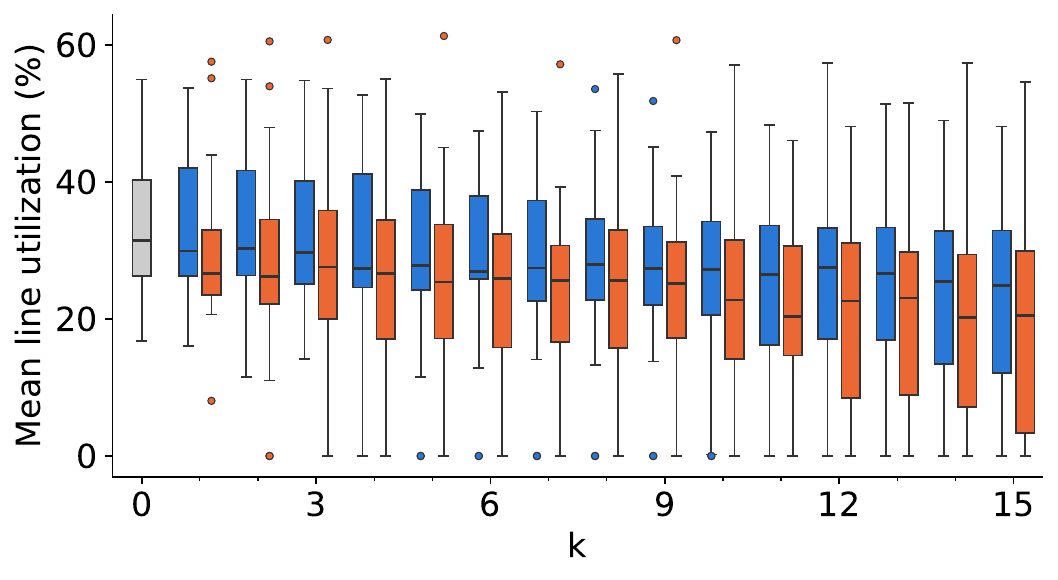}}
  \caption{Mean congestion in the network after $\budget$ or $\budget+\budget[']$ failures. Utilization of the network falls as $\budget$ grows; this occurs with smaller $\budget$ in the $\Nkkp$ model than the standard $\Nk$ model.}
  \label{fig:congestion-comparison}
\end{figure}

Our next experiment examines the state of the network that remains after failures and its load shed.
For $\budget$ from $1$ to $15$, we fail the lines selected by the standard $\Nk$ model ($\budget$ lines) and by the decoupled $\Nkkp$ model ($\budget + \budget[']$ lines), and report the utilization of each surviving line, i.e., its power flow as a fraction of its thermal rating.
The failed lines are excluded from this average, since they carry no flow and would otherwise be counted as uncongested.
Figure~\ref{fig:congestion-comparison} shows the average line utilization across the case studies as $\budget$ grows, with the fully operational network plotted at $\budget{\,=\,}0$.
Before failure, the benchmark networks are loaded to an average of $27.2\%$ of their thermal ratings under typical conditions and $33.8\%$ under heavy load; individual networks range from $10.1\%$ to $46.5\%$ and from $16.8\%$ to $54.9\%$, respectively.
As $\budget$ grows, the utilization of the network falls: at $\budget {\,=\,} 15$, the average utilization of the surviving lines is $19.9\%$ for the standard $\Nk$ model against $16.3\%$ for the decoupled $\Nkkp$ model on the regular cases, and $22.0\%$ against $19.0\%$ on the API cases.
This decline is faster under the $\Nkkp$ model because it sheds more load at the same budget: to cause the load shed that the $\Nkkp$ model causes at $\budget$, the standard model requires a budget larger by an average of $1.2$ on the regular cases and $2.6$ on the API cases.
Overall, this shows that the $\budget[']$ response shapes which components fail and how much load is lost, as well as the operating envelope of the remaining network.

\subsection{Utilizing more complex response models}
\label{sec:decoupled-approach-interoperability}

One of the contributions of the decoupling is its interoperability with other, more complex methods for computing the $\budget[']$ failures. 
Our next experiment demonstrates one such example using PowerSimulationDynamics~\cite{lara2023powersimulationsdynamicsjl}, a simulator capturing power system dynamics for a grid network, to compute the failures $\xlinevec[']$ caused by an initial failure $\xlinevec$. The standard defender is then evaluated for the $\xlinevec + \xlinevec[']$ combined failures. 
We assume that initial lines that are failed cannot disconnect the network (this is a limitation imposed by the simulation library). In our implementation, when an attack would island the network, we add a no-good cut to the attacker's set of constraints to exclude that $\xlinevec$. However, we note that this limitation prevents values of $\budget$ greater than $2$ from being sufficiently explored. This experiment is also only conducted on modified Case 240 system that includes additional parameters for dynamic generator and inverter devices that are necessary for modeling the dynamics (PDEs). 

\begin{table}[!ht]
\caption{Results from Simulation-based Response Model}
\label{tab:simulation-model-comparison}
\begin{tabular}{ccccc}
\hline
$\budget$ & & Load Shed (p.u.) & Runtime (s) & Iterations \\
\hline
\multirow{2}{*}{1} & 
    Algorithm~\ref{alg:decoupled-response-model} & 381.66 & 2.89   & 44 \\
&   Ref~\cite{lara2023powersimulationsdynamicsjl}             & 431.29 & 106.95 & 33 \\
\hline
\multirow{2}{*}{2} & 
    Algorithm~\ref{alg:decoupled-response-model} & 387.77 & 245.21  & 11325 \\
&   Ref~\cite{lara2023powersimulationsdynamicsjl}             & 439.38 & 6558.69 & 9814 \\
\hline
\end{tabular}
\end{table}

Table~\ref{tab:simulation-model-comparison} reports the load shed, run time, and number of iterations taken to converge to a solution using the two different response models. 
Our results show that the simulation-based response yields higher load shed with a larger number of lines that fail during the immediate automated system response. This is seen for both $\budget{=}1$ and $\budget{=}2$. 
However, this comes at the cost of a largely increased runtime, since attacks that island the network are disallowed.
Another interesting observation is that the run time and number of iterations also become less correlated; many iterations may be spent cutting off infeasible choices of $\xlinevec$ without evaluating the defender model. 
These results indicate that by simulating the transience of the network in more detail, it is possible to find larger, different sets of $\budget[']$ failures. 
This highlights a key result of our paper: \emph{the decoupled approach enables interoperability with arbitrarily complex response models while preserving the optimality guarantees of the attacker-defender optimization. 
In practice, this allows flexibility in studying and capturing realistic grid failures.}

\section{Conclusions and Future Directions}
\label{sec:conclusion}

In this paper, we presented a decoupled approach for worst-case contingency analysis that extends the standard $\Nk$ interdiction framework to an $\Nkkp$ setting by explicitly modeling the grid’s immediate response. 
Our key insight is that, because primal cuts are constructed entirely in the space of the initial set of failures $\xlinevec$, the computation of induced failures  $\xlinevec [']$ can be pulled out of the defender and treated as a deterministic black box, all without altering cut construction or compromising optimality. 
We formalized this insight and integrated it into a primal cut algorithm, which enables efficient evaluation of the attacker, defender, and the response model in between. 
We also showed that the defender operating on the combined failures $\xlinevec + \xlinevec[']$ produces the same cut coefficients as the integrated model. 
Our experiments show that across benchmark networks, accounting for induced failures substantially increases identified worst-case load shed relative to the standard $\Nk$ model. However, the decoupled formulation scales more reliably than an integrated alternative. 
We further demonstrated interoperability by substituting a high-fidelity dynamics simulator as the response module, which revealed larger induced failures (and higher load shed). 
Future work will examine interdependencies and apply such analysis to other related infrastructures such as transportation and water. 
We also plan to study the impacts of different geo-correlated disaster settings, and analyze the potential failures they cause for the electric grid.

\section*{Acknowledgments}
This research was supported by the UC National Lab Fees Research Program Grant No. L22GF4561, and NSF Grants No. 1952247 and 2420846.
The work was also partially funded by Los Alamos National Laboratory's Directed Research and Development project, ``Artificial Intelligence for Mission (ArtIMis)'' and the U.S. Department of Energy Office of Electricity Project ``North American Energy Resilience Model'' under U.S. DOE Contract No. DE-AC52-06NA25396.
The authors would like to acknowledge Manuel Garcia from the Los Alamos National Laboratory for his participation and contributions in discussions. 

\bibliographystyle{IEEEtran}
\bibliography{bibliography-short}



\end{document}